\documentclass[aps,prb,twocolumn,groupedaddress,showkeys,showpacs]{revtex4}

\usepackage{multirow,amssymb,amsbsy,amsmath}
\usepackage{epsfig}
\usepackage{graphicx}
\usepackage{epstopdf}

\newcommand {\mofe} {\{$\textrm{Mo}_{72}\textrm{Fe}_{30}$\}}
\newcommand {\mocr} {\{$\textrm{Mo}_{72}\textrm{Cr}_{30}$\}}

\begin{document}

\title{Multiple nearest-neighbor exchange model for the frustrated magnetic molecules \mofe\, and \mocr}

\author{Christian Schr\"oder}
\email{christian.schroeder@fh-bielefeld.de}
\affiliation{Department of Electrical Engineering and Computer
  Science, University of Applied Sciences Bielefeld, D-33602 Bielefeld, Germany
  \& Ames Laboratory, Ames, Iowa 50011, USA}

\author{Ruslan Prozorov}
\author{Paul K\"ogerler}
\author{Matthew D. Vannette}
\author{Xikui Fang}
\author{Marshall Luban}
\affiliation{Ames Laboratory \& Department of Physics and Astronomy,
Iowa State University, Ames, Iowa 50011, USA}

\author{Akira Matsuo}
\author{Koichi Kindo}
\affiliation{Institute for Solid State Physics, University of Tokyo, Kashiwanoha 5-1-5, Kashiwa, Chiba 277-8581, Japan}

\author{Achim M\"uller}
\author{Ana Maria Todea}

\affiliation{Fakult\"at f\"ur Chemie, Universit\"at Bielefeld, D-33501 Bielefeld, Germany}

\date{\today}

\begin{abstract} 
Our measurements of the differential susceptibility $\partial M/\partial H$ 
of the frustrated magnetic molecules {\{}Mo$_{72}$Fe$_{30}${\}} 
and {\{}Mo$_{72}$Cr$_{30}${\}} reveal a pronounced dependence on magnetic 
field ($H$) and temperature ($T$) in the low $H$ - low $T$ regime, contrary to the 
predictions of existing models. Excellent agreement with experiment is 
achieved upon formulating a nearest-neighbor classical Heisenberg model 
where the 60 nearest-neighbor exchange interactions in each molecule, rather 
than being identical as has been assumed heretofore, are described by a 
two-parameter rectangular probability distribution of values of the exchange constant. 
We suggest that the probability distribution provides a convenient 
phenomenological platform for summarizing the combined effects of multiple 
microscopic mechanisms that disrupt the idealized picture of a Heisenberg 
model based on a single value of the nearest-neighbor exchange constant.
\end{abstract} 
\maketitle

%%%%%%%%%%%%%%%%%%%%%%%%%%%%%%%%%%%%%%%%%%%%%%%%%%%%%%%%%%%%%%%%%%%%%%%%
\section{\label{sec:intro}Introduction}
In recent years there has been extensive research on magnetic molecules as 
these are novel, realizable systems for exploring magnetic phenomena in 
low-dimensional magnetic materials.\cite{1,2,3,4,5} Among these diverse systems 
the pair of Keplerate structural type magnetic molecules abbreviated as 
{\{}Mo$_{72}$Fe$_{30}${\}} and {\{}Mo$_{72}$Cr$_{30}${\}}, each hosting a 
highly symmetric array of 30 exchange-coupled magnetic ions (``spin 
centers''), serve as highly attractive targets for the investigation of 
frustrated magnetic systems. In these molecules\cite{6,7} the magnetic ions 
Fe$^{\mbox{\footnotesize{III}}}$ (spin $s = 5/2$) and Cr$^{\mbox{\footnotesize{III}}}$ (spin $s$ = 3/2) occupy the 30 
symmetric sites of an icosidodecahedron, a closed spherical structure consisting of 20 
corner-sharing triangles arranged around 12 pentagons (diamagnetic polyoxomolybdate fragments). This is
a zero-dimensional analogue of the planar kagome lattice that is composed of corner-sharing 
triangles arranged around hexagons. A useful theoretical framework that has 
been employed\cite{8,9,10} in studying these magnetic molecules is based on an 
isotropic Heisenberg model, where each magnetic ion is coupled via 
intra-molecular isotropic antiferromagnetic exchange to its four nearest 
neighbor magnetic ions, and all of the 60 intra-molecular exchange 
interactions are of equal strength (henceforth, ``single-$J$ model'').\cite{11} 
Unfortunately the quantum Heisenberg model of the two magnetic molecules is 
intractable using either analytical or matrix diagonalization methods. 
Nevertheless, the nearest-neighbor exchange constant for each molecule has 
been established by comparing experimental data above 30~K for the 
temperature-dependent zero-field susceptibility with data obtained by 
simulational methods using the quantum\cite{7} and classical\cite{9} Monte Carlo 
methods. For temperatures below about 30~K, where the quantum Monte Carlo 
method proves to be ineffective for the two magnetic molecules due to frustration 
effects,\cite{12} the classical Heisenberg model is at present the only 
practical platform for establishing the dependence of the magnetization $M(H,T)$ on 
external magnetic field $H$ and temperature $T$. A rigorous analytical result\cite{8} for the 
classical, nearest-neighbor, single-$J$ Heisenberg model states that, in the zero temperature 
limit, $M$ is linear in $H$ until $M$ saturates (saturation fields $H_{s}$ = 17.7~T and 
60.0~T for {\{}Mo$_{72}$Fe$_{30}${\}} and {\{}Mo$_{72}$Cr$_{30}${\}}, 
respectively). The practical relevance of the classical Heisenberg model in 
describing these magnetic molecules even at low temperatures was strikingly 
demonstrated in an earlier experiment\cite{9} on {\{}Mo$_{72}$Fe$_{30}${\}} at 
0.4~K, showing an overall linear dependence of $M$ on $H$ and its saturation at 
approximately 17.7~T. The ground state envisaged by the classical single-$J$ 
model is characterized by high-symmetry spin frustration. In particular, for 
$H=0$, in the ground state the spins are coplanar with an angular separation 
of 120$^\circ$ between the orientations of nearest-neighbor spins. On 
increasing the external field $H$ the spin vectors gradually tilt towards the 
field vector, until full alignment is achieved when $H=H_s$, while their 
projections in the plane perpendicular to the field vector retain the 
120$^\circ$ pattern for nearest-neighbor spins.\cite{8} 
It would be reasonable to expect that $M(H,T)$ is an analytic function of its 
variables and thus $\partial M/\partial H\approx M_s/H_s$ for $H < H_s$ and 
$k_B T \ll J_0$, essentially independent of \textit{both} $H$ and $T$ in these 
intervals, where $J_0$ is the exchange constant of the single-$J$ classical Heisenberg model.\cite{11} 
Indeed our classical Monte Carlo calculations confirm this behavior.

As shown here, new and crucial features of $M(H,T)$ become accessible upon 
examining the differential susceptibility, $\partial M/\partial H$, as a 
function of $H$ and $T$. What emerge are significant conflicts, spelled out in Secs.~\ref{sec:Exp:A} and \ref{sec:Exp:B} 
between the results of our measurements and a theory based on the classical 
single-$J$ model as concerns both the $T$ and $H$ dependence of $\partial M/\partial 
H$ below 5~K. However, and this is the central idea of this 
paper, agreement with experiment is achieved upon adopting a refined 
classical Heisenberg model of {\{}Mo$_{72}$Fe$_{30}${\}} and 
{\{}Mo$_{72}$Cr$_{30}${\}}, where we drop the assumption of a 
single, common value for the nearest-neighbor exchange constant, using 
instead a rectangular probability distribution\cite{15} for the 60 nearest-neighbor 
intra-molecular interactions. 
%--------------------------- figure ----------------------------------------
\begin{figure}
[phb]
\begin{center}
\includegraphics[
width=8.5cm
]%
{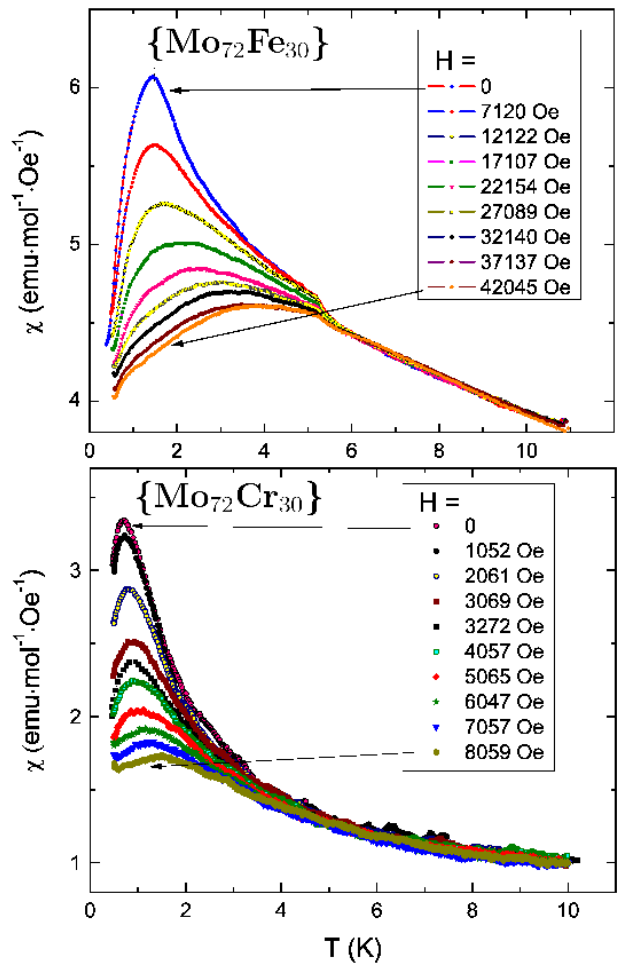}%
\caption{Temperature dependence of the measured differential susceptibility 
$\partial M/\partial H$ for {\{}Mo$_{72}$Fe$_{30}${\}} (upper panel) and for {\{}Mo$_{72}$Cr$_{30}${\}} (lower panel). Values of $H$ are listed in the 
legends.}%
\label{fig1}%
\end{center}
\end{figure}
%--------------------------- figure ----------------------------------------

The layout of this paper is as follows. In Sec.~\ref{sec:Exp:A} we present our 
experimental results for $\partial M/\partial H$ as a function of $T$ in the 
low-temperature range for {\{}Mo$_{72}$Fe$_{30}${\}} and 
{\{}Mo$_{72}$Cr$_{30}${\}} for several low field values. In Sec.~\ref{sec:Exp:B} we 
present our experimental results for $\partial M/\partial H$ as a function of $H$. In Sec.~\ref{sec:iso} we show that the model system of a classical isosceles spin triangle exhibits the same qualitative features as the 
results of our experiments summarized in Sec.~\ref{sec:Exp:A}. In Sec.~\ref{sec:mul}, using the 
classical Heisenberg model of the icosidodecahedron and Monte Carlo 
simulational methods, we show that excellent agreement between a model calculation based on a two-parameter distribution of exchange constants and our experimental data of Sec.~\ref{sec:Exp} can be achieved. Finally, in Sec.~\ref{sec:sum} we summarize 
our findings, discuss the broader implications of our results, and identify 
several open questions. 
\section{\label{sec:Exp}EXPERIMENT}
\subsection{\label{sec:Exp:A}$\partial M/\partial H$\textbf{\textit{ 
}}\textbf{versus}\textbf{\textit{ T}}}
Measurements of $\partial M/\partial H$ versus $T$ were performed at Ames Laboratory on 
polycrystalline samples of {\{}Mo$_{72}$Fe$_{30}${\}} and 
{\{}Mo$_{72}$Cr$_{30}${\}} (see experimental sections of Refs.~\onlinecite{6} and \onlinecite{7}) prepared by employing optimized synthesis methods 
and re-crystallization steps to minimize paramagnetic impurities. 
Our data was obtained by using a self-resonating LC circuit driven by a 
tunnel diode.\cite{18,19} Briefly, a tank circuit consisting of a small coil 
of inductance $L_0$ and a capacitor, $C$, is kept at constant temperature of $5\pm0.005$~K. The sample is mounted on a sapphire holder, which is 
inserted into the coil without making contact. In the absence of the sample, 
the circuit resonates at the resonant frequency $2\pi f_0 = 1/\sqrt{LC}$.
When a sample with susceptibility $\chi$ is inserted into the coil, the 
resonant frequency changes from $f_{0}$ to $f(\chi )$ due to the change of the coil 
inductance, $L=d\Phi /dI$, where $\Phi$ is the total magnetic flux through 
the coil and $I$ is the current in the coil. This current generates a magnetic 
field of about 20~mOe. If the magnetic perturbation due to the sample is 
small, the shift of the resonant frequency is given by
\[
\frac{\Delta f}{f_0 }\approx \frac{\Delta L\left( \chi \right)}{2L_0 }=\frac{4\pi 
\chi }{(1-N)}\frac{V}{2V_0 }\quad .
\]
Here $N$ is the demagnetization factor, $\Delta L\left(  \chi\right)  =\left| L\left(  \chi\right)
-L_{0} \right| \ll L_{0}$ is the change of the 
coil inductance, $V$ is the sample volume and $V_0$ is volume of the coil, and 
$\chi$ is the dynamic magnetic susceptibility at our typical resonant frequency of 
10~MHz. This frequency is still much lower than characteristic frequencies 
and the response can be considered as static, i.e., $\chi$ can be identified with 
$\partial M/\partial H$. Measured frequency shifts for the samples described 
in this work are of the order of 1 to 10~Hz, whereas the experimental 
resolution of the setup is about 0.01~Hz which corresponds to a smallest 
detectable magnetic moment of about 5~picoemu. For easy comparison with 
other experiments, the measured frequency shift is proportional to the 
change in the total magnetic moment of the sample at a given frequency, 
$\Delta M=C\Delta f$, where $C$ is a calibration constant.

In Fig.~\ref{fig1} we show our experimental data for $\partial M/\partial H$ 
versus $T$ for {\{}Mo$_{72}$Fe$_{30}${\}} and 
{\{}Mo$_{72}$Cr$_{30}${\}}, respectively. The calibration of the 
experimental data was achieved by matching the effective magnetic moment 
inferred from the measured frequency shift to low-field (3 - 50000 Gs) ac (10 to 1000 Hz) and dc susceptibility 
measurements performed on a Quantum Design MPMS, which show that at high 
temperatures dynamic effects can be neglected. The assumption of static 
behavior is supported by the fact that all curves collapse onto each other 
for $T > 5$ K. The most striking feature of these data is the very strong 
dependence of $\partial M/\partial H$ on $H$ and $T$. As remarked in Sec.~\ref{sec:intro}, this 
behavior is contrary to that predicted by the single-$J$ model, 
specifically $M\approx H\chi_0 (k_B T/J_0)\approx H\chi_0(0)$, that is, 
$\partial M/\partial H$ is independent of $H$ and $T$ in the regime $H \ll H_s ,\;k_B 
T \ll J_0$.
\subsection{\label{sec:Exp:B}$\partial M/\partial H$\textbf{ versus }\textbf{\textit{H}}}
In Fig.~\ref{fig3} we show our experimental data for $\partial M/\partial H$ versus $H$ for both {\{}Mo$_{72}$Fe$_{30}${\}} (upper panel) and {\{}Mo$_{72}$Cr$_{30}${\}} (lower panel).  These data sets were obtained  by measuring the magnetization using pulsed magnetic fields (typical sweep rate 15000 $T/s$) with a standard inductive method at facilities of Okayama University, Tohoku University, and University of Tokyo.  
Also shown are the corresponding classical Monte Carlo simulational 
results for the single-$J$ model (dashed curve) for the indicated temperatures.
%--------------------------- figure ----------------------------------------
\begin{figure}
[phb]
\begin{center}
\includegraphics[
width=8cm]%
{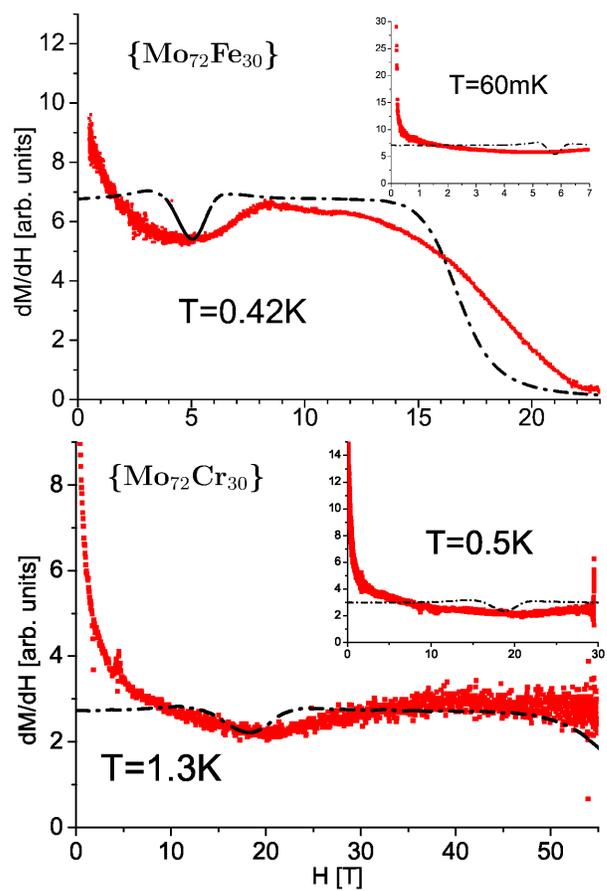}%
\caption{Magnetic field dependence of the measured differential 
susceptibility $\partial M/\partial H$, shown in red, for {\{}Mo$_{72}$Fe$_{30}${\}} for $T$ = 0.42~K and 60~mK (upper panel, inset) and for {\{}Mo$_{72}$Cr$_{30}${\}} for $T$ = 1.3~K and 0.5~K (lower panel, inset). The dashed 
curves are the results of the single-$J$ model for these temperatures.}%
\label{fig2}%
\end{center}
\end{figure}
%--------------------------- figure ----------------------------------------

The inadequacy of the single-$J$ model is striking in that the simulational 
data differ from the experimental data in four important ways. \textit{First}, the 
experimental data exhibit a steep rise for decreasing low fields, and the 
lower the temperature the steeper the rise. This is consistent with our 
experimental findings in Sec.~\ref{sec:Exp:A} for the $T$- dependence of $\partial 
M/\partial H$. These features are entirely absent for the single-$J$ model; in 
particular, $\partial M/\partial H$ is essentially independent of field in 
the low-field regime. \textit{Second}, the local minimum in the experimental data is 
significantly broader than that predicted by the single-$J$ model. \textit{Third}, according 
to the single-$J$ model a local minimum in $\partial M/\partial H$
versus $H $emerges at $T$ = 0 K at a field $H = H_{s}$/3 (approx. 6~T and 20~T for 
{\{}Mo$_{72}$Fe$_{30}${\}} and {\{}Cr$_{72}$Fe$_{30}${\}}, respectively) and 
it is enhanced with increasing $T$.\cite{17} By contrast, our experimental data 
for fields in the vicinity of $H_{s}$/3 differ insignificantly with 
temperature. \textit{Fourth}, for {\{}Mo$_{72}$Fe$_{30}${\}} the experimental data for 0.42~ 
K shows a decrease with increasing field above 10~T, quite distinct from the 
pattern of the single-$J$ model.
\section{\label{sec:iso}CLASSICAL ISOSCELES SPIN TRIANGLES}
In this Section we give a qualitative explanation for our experimental 
findings in Sec.~\ref{sec:Exp:A}. We suggest that the strong sensitivity of the 
differential susceptibility on $H$ and $T$ reflects a non-analytic dependence of 
the magnetization on these variables, a characteristic already exhibited by independent classical isosceles spin triangles. The Hamiltonian of a 
spin triangle is given in footnote~\onlinecite{16} in terms of dimensionless quantities. In particular, the three pair-wise 
interactions are described by two different antiferromagnetic exchange 
constants (positive values of $J$ and ${J}'$) and we consider both cases, 
$J'<J$ and $J'>J$. 

For the equilateral spin triangle ($J'=J=J_0)$ at $T = 0$~K the magnetic moment 
per triangle, $M(H,0)$, is a continuous, linear function of $H$ and, in 
particular, it vanishes for $H \to 0$. Specifically, $M(H,0)=3 H/H_s$ for 
$-H_s<H<H_s$, where $H_s =3J_0$ is the saturation field. 

Analytical calculation of $M(H,0)$ for the corresponding isosceles spin 
triangle is a non-trivial task for general values of $J/J'$ as it is 
necessary to carefully identify the configuration of least energy for 
arbitrary values of $H$. The final results are as follows: For the case $J'>J$: 
$M(H>0,0)=1-J/J'+(2+J/J')H/H_s$ for $0<H<H_s $, where $H_s =2J'+J$ is 
the saturation field. For the case $J'<J<2J'$: 
$M(H>0,0)=J/J'-1+H/J'$ for $0<H<2J'-J$; $M(H,0) = 1$ for $2J'-J<H<J$; and 
$M(H,0)=H/J$ for $J<H<3J=H_s$. $M(H,0)$ is subject 
to the antisymmetry property $M(-H,0)=-M(H,0)$. The discontinuous behavior for $H\to 0$ is 
of particular importance in both cases. Although this discontinuity occurs 
only at 0~K, it is the crucial mechanism responsible for the great 
sensitivity of $M(H,T)$ in the low $H$ - low $T$ regime.
%--------------------------- figure ----------------------------------------
\begin{figure}
[phb]
\begin{center}
\includegraphics[
width=10cm
]%
{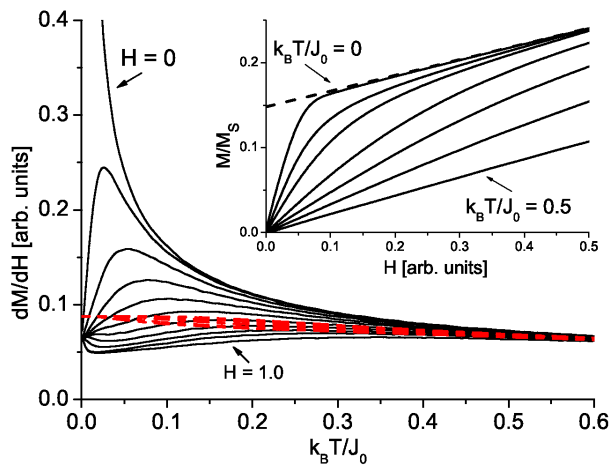}%
\caption{Temperature dependence of the differential susceptibility $\partial 
M/\partial H$ for a classical isosceles spin triangle (black curves) with $J/k_B = 
1$~K and $J'/k_B = 1.8$~K for field values $H$ = 0, 0.1, {\ldots}, 1. 
Inset: $M/M_s$ versus $H$ for the values $k_B T/J_0 = 0, 0.001, 0.01, 0.02, 
0.05, 0.1, 0.2, 0.5$. Results for the corresponding classical equilateral 
spin triangle ($J_0/k_B=1.267$)~K are given by the red dashed curves.}%
\label{fig3}%
\end{center}
\end{figure}
%--------------------------- figure ----------------------------------------

The form of $M(T,H)$ for finite $T$ can, in principle, be derived for this model 
system by analytical methods, however these calculations are substantially 
more intricate than for the classical equilateral spin triangle (see Sec.~\ref{sec:Exp} 
B of O. Ciftja \textit{et al.} listed in Ref.~\onlinecite{11}). For practical purposes, the simplest 
procedure is to use the classical Monte Carlo method for convenient 
numerical choices of $J'/J$. In Fig.~\ref{fig3} we display our results for 
the choices $J/k_B=1$~K, $J'/k_B = 1.8$~K. The 
most important feature to note is that while $M$ is indeed a continuous function 
of $H$ for any nonzero $T$, the quantity $\partial M/\partial H$, provided in the 
main portion of Fig.~\ref{fig3}, exhibits strong temperature dependence for 
weak magnetic fields. Indeed, the curves fan with increasing $H$ in a manner 
that is strikingly similar to that shown in Fig.~\ref{fig1}. 
Similar behavior occurs for cases where $J'<J$. By contrast, for the corresponding classical equilateral spin triangle $\partial M/\partial H$
is virtually independent of $T$, as is seen in Fig.~\ref{fig3}.
\section{\label{sec:mul}MULTIPLE NEAREST-NEIGHBOR COUPLINGS}
The $T$ and $H$ dependence of $\partial M/\partial H$ for the classical isosceles 
spin triangle considered in the previous section is remarkably similar to 
that of our experimental data in Sec.~\ref{sec:Exp}. However, to achieve a more 
realistic model, in the following we assume that the 60 nearest-neighbor 
exchange interactions between magnetic ions in a given molecule 
are characterized by a probability distribution with two adjustable width 
parameters. In the following Section we rationalize the use of a probability distribution as a convenient way for summarizing the combined effects of multiple microscopic mechanisms that disrupt the use of an idealized, single-$J$ model.

We simulate each of {\{}Mo$_{72}$Fe$_{30}${\}} and 
{\{}Mo$_{72}$Cr$_{30}${\}} by considering an ensemble of up to 100 
independent systems, for a total of 60 couplings per 
system. We assign values of the 6000 exchange 
constants using a random number generator according to the following 
rules: 1) The average value, $J_{0n}$, of the classical exchange constant 
(in units of Boltzmann's constant) for the $n$th system is allowed to assume 
any value in the interval $\left( (1-\tau )J_0 ,\;(1+\tau )J_0  \right)$ 
with equal probability, where $J_0$ is chosen as 13.74~K for 
{\{}Mo$_{72}$Fe$_{30}${\}} and 32.63~K for {\{}Mo$_{72}$Cr$_{30}${\}} as 
determined by high-temperature susceptibility measurements\cite{11}; 2) For the $n$th 
system, the individual values of the classical exchange constant are allowed 
to assume any value in the interval $\left( (1-\rho )J_{0n} ,\;(1+\rho 
)J_{0n}  \right)$ with equal probability. For each molecule the two 
parameters $\tau ,\;\rho $ characterizing these rectangular probability distributions were 
determined so as to provide an optimal fit with our experimental data for 
$\partial M/\partial H$ versus $H$. 

%--------------------------- figure ----------------------------------------
\begin{figure}
[phb]
\begin{center}
\includegraphics[
width=8cm
]%
{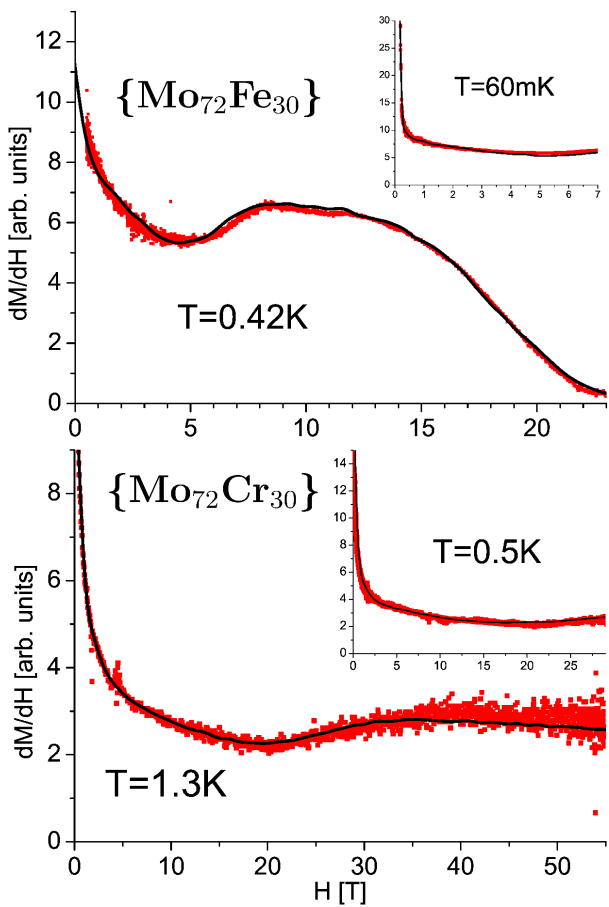}%
\caption{Measured differential susceptibility $\partial M/\partial H$ versus 
$H$, shown in red, for {\{}Mo$_{72}$Fe$_{30}${\}} for $T$ = 0.42~K and 60~mK (inset) and for {\{}Mo$_{72}$Cr$_{30}${\}} for $T$ = 1.3~K (inset: 0.5~K) and simulational results (solid black curve) using a multiple-$J$ model for 
the optimal choice of the probability distribution parameters as given in the 
text.}%
\label{fig4}%
\end{center}
\end{figure}
%--------------------------- figure ----------------------------------------
In Fig.~\ref{fig4} we present our results for $\partial M/\partial H$ versus $H$ for 
{\{}Mo$_{72}$Fe$_{30}${\}} and {\{}Mo$_{72}$Cr$_{30}${\}}. Note the excellent agreement between the experimental 
data and the simulational results obtained using our multiple-$J$ model (solid 
curve). In the case of {\{}Mo$_{72}$Fe$_{30}${\}}, the optimal choices of 
the parameters $\tau ,\;\rho $ were $\tau  = 0.15$ and $\rho = 0.40$, whereas for 
{\{}Mo$_{72}$Cr$_{30}${\}} these were $\tau = 0$ and $\rho  = 0.5$. 

The motivation for using two distributions as described above is the 
following. One can attribute the influence of the two distributions, characterized by $\tau$ and $\rho$, 
respectively, to complementary effects which only \textit{in combination} lead to the observed properties 
of both molecules.
The value of $\tau$ controls the variation in the values of the 
mean exchange constant \textit{per molecule}, which leads to variations in the value of the 
saturation field $H_s$ and hence in the value of the minimum in $\partial M/\partial H$ versus $H$ 
at $H_s/3$. By averaging over those variations one finds that $\partial M/\partial H$ versus $H$ 
starts to decrease at a much lower value of $H$ than predicted by the single-$J$ model and 
\textit{simultaneously} finds that the minimum at $H_s/3$ is broadened as observed. 
However, using this distribution alone one cannot explain the observed strong $H$ dependence 
of $\partial M/\partial H$ in the low $T$ - low $H$ regime, because each molecule is still characterized by a single exchange constant. Introducing a second distribution, characterized by the 
parameter $\rho$, leads to a variation in the values of the exchange constant 
\textit{within a molecule} with the effect that the corner-sharing spin triangles are of the scalene-type rather than equilateral-type. This gives rise to the non-analytic 
behavior of the magnetization at $H=0$ for 0~K and hence to the characteristic effects we have found in the low $H$ - low $T$ regime. In our simulations we have studied a 
very large range of choices of parameter pairs. The optimal choice for 
{\{}Mo$_{72}$Fe$_{30}${\}} can be narrowed to $\tau = 0.15\pm0.02$ and $\rho = 
0.40\pm0.02$. For {\{}Mo$_{72}$Cr$_{30}${\}} we find $\rho = 0.50\pm0.02$, however $\tau $ can be chosen in the range 0 to 0.2 without any observable 
effect. This is due to the fact that for {\{}Mo$_{72}$Cr$_{30}${\}} 
magnetization measurements above the saturation field ($H_s$ = 60~T) are 
not achievable at the present time. In the case of 
{\{}Mo$_{72}$Fe$_{30}${\}}, for which $H_s$ = 17.7~T, the availability of 
magnetization data above the saturation field allows for a greatly reduced 
uncertainty in the value of $\tau$.
%--------------------------- figure ----------------------------------------
\begin{figure}
[phb]
\begin{center}
\includegraphics[
width=8cm
]%
{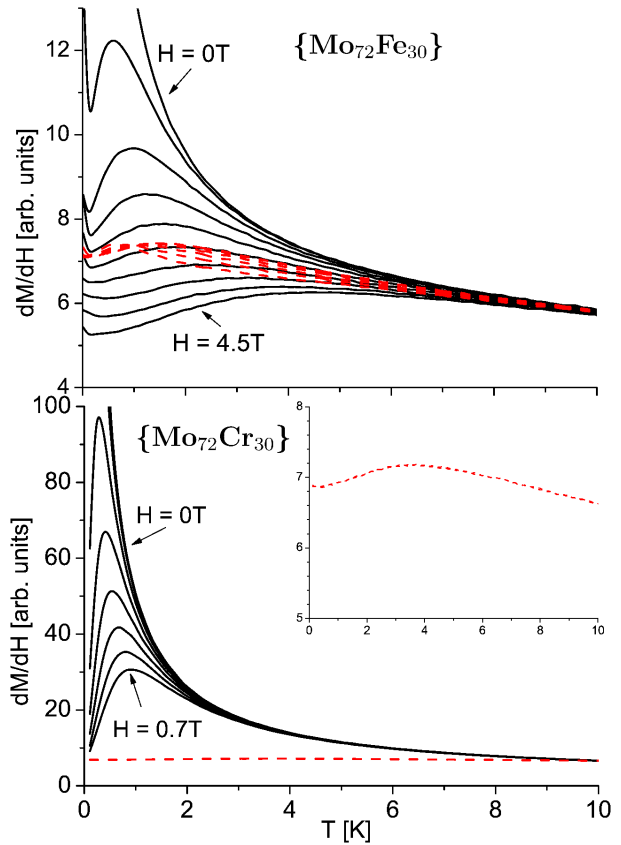}%
\caption{Simulational results for $\partial M/\partial H$ versus $T$ based on the multiple-$J$ model using the optimal distribution
parameters for {\{}Mo$_{72}$Fe$_{30}${\}} and for {\{}Mo$_{72}$Cr$_{30}${\}} as given in the text. Results for the corresponding single-$J$ model 
calculations are shown as red dashed curves.}%
\label{fig5}%
\end{center}
\end{figure}
%--------------------------- figure ----------------------------------------

Shown in Fig.~\ref{fig5} are our simulational results for $\partial 
M/\partial H$ versus $T$ for several different values of $H$ using the probability 
distribution with optimal parameters appropriate for 
{\{}Mo$_{72}$Fe$_{30}${\}} and {\{}Mo$_{72}$Cr$_{30}${\}}. These results 
(black curves) are strikingly similar to the experimental curves seen in 
Fig.~\ref{fig1}; the corresponding curves (shown in red) 
for the single-$J$ model, for the same choice of the mean value $J_0$, are 
essentially indistinguishable from one another. This again strongly supports 
the existence of the multiple-$J$ scenario. Note that, in the experiment as 
well as in the simulations, with increasing temperature the curves for 
different field values rapidly converge and become indistinguishable from 
one another. Also, for increasing temperature, the results for the multiple-$J$ 
model merge with those of the single-$J$ model, as expected, since the average 
exchange constant across the ensemble, $J_0$, is chosen to equal to the 
exchange constant of the single-$J$ model. Finally, it remains to be seen 
whether the sharp rise in the curves of the upper panel of Fig.~\ref{fig5} below 200~mK is an 
experimental feature in {\{}Mo$_{72}$Fe$_{30}${\}} or merely an artifact of 
the multiple-$J$ model based on a rectangular probability distribution.
\section{\label{sec:sum}SUMMARY AND DISCUSSION}
In this article, we have presented our experimental data for the 
differential susceptibility of the pair of magnetic molecules 
{\{}Mo$_{72}$Fe$_{30}${\}} and {\{}Mo$_{72}$Cr$_{30}${\}} as a function of 
magnetic field and temperature. Below 5~K these data are strikingly different from 
what can be provided using a classical Heisenberg model with a single value 
of the nearest-neighbor exchange constant (single-$J$ model). We have achieved 
excellent agreement with our experimental data upon adopting a classical 
Heisenberg model where the 60 nearest-neighbor interactions are not 
identical; instead, the values of the exchange constants are described by a 
two-parameter probability distribution with a mean value as determined 
from experimental $\partial M/\partial H$ data above 30~K using the single-$J$ model.
Above 5~K the single-$J$ model provides a satisfactory description of each molecule.

Since the icosidodecahedron structure consists of corner-sharing triangles, 
it is not surprising that the Heisenberg model of \textit{independent} classical isosceles spin 
triangles provides a simple yet instructive model in that it exhibits the main qualitative
features of our experimental data. We note here that a similar approach has been employed successfully for various two-dimensional spin systems on triangular lattices\cite{20}. For example, in the case of manganese tricyanomethanide the so-called `row model' based on \textit{connected} isosceles triangles provides the explanation for an unusual magnetic-field dependence of the spin ordering\cite{24}. In the context of independent classical isosceles spin triangles one can figuratively describe the effect of multiple exchange constants as modifying the spin frustration from the standard $120^\circ$ 
angular separation between spin vectors of the equilateral spin triangle. 
The operational consequence is that the magnetization, for $T=0$, of an isosceles spin 
triangle is a non-analytic function of magnetic field for $H=0$, 
and this is manifested in $\partial M/\partial H$ being a highly sensitive 
function of its arguments for small $H$ and $T$.

The existence of a distribution of nearest-neighbor exchange constants can 
be expected to be responsible for a significant lifting of degeneracies of 
magnetic energy levels. To be specific, the quantum rotational band 
model\cite{10}, which is a solvable alternative to the nearest-neighbor 
single-$J$ quantum Heisenberg model, predicts a discrete spectrum of energy 
levels, many of which have a very high degeneracy due to large multiplicity 
factors. Perturbing this model Hamiltonian by using a distribution of 
$J$-values would remove a major fraction of these degeneracies. The lifting of 
level degeneracies could provide a reasonable explanation for three 
long-standing puzzling issues concerning these magnetic molecules:
The first issue is the very broad peak (maximum at 0.6~meV) that has been 
observed by inelastic neutron scattering on {\{}Mo$_{72}$Fe$_{30}${\}} at 65~mK. In order to qualitatively reproduce the observed peak using the 
rotational band model, it was necessary in Ref.~\onlinecite{21} to perform the 
calculations upon assigning a large energy width (0.3~meV) for the 
individual energy levels. We suggest that the source of this large energy 
width might be the lifting of the majority of degeneracies associated with a 
single-$J$ model. 

A second important consequence of the splitting of highly degenerate levels 
would be that the molecules could exhibit classical characteristics down to 
very low temperatures. This would provide a very reasonable explanation for 
the surprising fact that our simulational results based on the \textit{classical} Heisenberg 
Hamiltonian are so successful in describing {\{}Mo$_{72}$Cr$_{30}${\}}, 
despite the fact that the Cr$^{\mbox{\footnotesize{III}}}$ ions have a small spin (3/2). Stated 
differently, with the lifting of degeneracies and the fanning out of energy 
levels the effective temperature for the crossover from classical to quantum 
behavior can be anticipated to be considerably lower than that expected \textit{a priori} for 
the single-$J$ model.

Third, the failure of efforts to observe magnetization steps, in 
measurements of magnetization versus $H$, in the mK temperature range in both {\{}Mo$_{72}$Fe$_{30}${\}} and 
{\{}Mo$_{72}$Cr$_{30}${\}} could also be attributed to the removal of degeneracies of the 
magnetic energy levels. The occurrence of magnetization steps at low 
temperatures is associated with the field-induced crossing of successive 
energy levels of the lowest rotational band. However, a discrete level 
associated with total spin quantum number $S$ has multiplicity $2S+1$ [total 
degeneracy $(2S+1)^2$]. If this degeneracy is lifted there will be a 
multitude of level crossings at slightly different field values and thus 
give rise to blurred effects down to lower temperatures than would otherwise 
be expected.

Given the finite-spin values of the Fe$^{\mbox{\footnotesize{III}}}$ and Cr$^{\mbox{\footnotesize{III}}}$ ions, is it 
possible to explain the present experimental findings based on a \textit{quantum} Heisenberg 
model that adopts a common, \textit{single} value of the exchange constant for all of the 
nearest-neighbor interactions? We strongly doubt that this is possible, for 
we have seen, albeit with a \textit{classical} Heisenberg model, that it is the spread in 
values of the nearest-neighbor exchange constant that fuels the sensitive 
dependence of $\partial M/\partial H$ on $T$, or equivalently the non-analytic 
behavior of magnetization on $H$ in the low $H$ - low $T$ regime.

Basing our simulations on a probability distribution for the 
nearest-neighbor exchange interaction has led to excellent agreement with 
the detailed features of our experimental data including the sensitive 
dependence of $\partial M/\partial H$ on $T$ and $H$. One can attribute the 
failure of the single-$J$ model to the combined effect of a large number of 
diverse perturbing mechanisms. The effects of impurities, variations in the 
exchange-coupling geometry, weak magnetic exchange interactions of 
more-distant neighbors, Dzyaloshinsky-Moriya and dipole-dipole interactions 
in these magnetic molecules are some of the many effects that are excluded when one 
uses an idealized single-$J$ model. On the other hand, it is at this stage an 
extremely difficult, essentially impossible task to realistically quantify 
the effects of the diverse mechanisms. A theoretical description based on a 
Heisenberg model where the nearest-neighbor exchange constant is chosen 
using a probability distribution provides a relatively simple, 
phenomenological platform for compromising between the need for microscopic 
realism versus practical limitations. Ultimately it is significant that a 
two-parameter probability description can actually provide the level of 
agreement that we have found. Finally, we remark that 
other choices of probability distributions can be expected to perform equally well.

As one example of the complications in assessing the plethora of perturbing 
mechanisms, we consider the variation in the intramolecular distances 
between nearest-neighbor magnetic ions. A geometric analysis utilizing 
existing low-temperature single crystal X-ray structure data for 
{\{}Mo$_{72}$Fe$_{30}${\}} molecules shows that the substructure of the 
magnetic ions is close to an ideal $I_{h}$-symmetric geometry. There is a 
standard deviation of 0.04525~{\AA} (0.70{\%}) and a maximum deviation of 
1.4{\%} from the average Fe-Fe distance of 6.4493~{\AA} for all 60 Fe-Fe 
nearest-neighbor distances. Besides the distances between the spin centers, 
geometric variations within the polyoxomolybdate exchange ligand are 
observed. In particular the O-Mo-O bond angle variations should affect the 
total orbital overlap and thus the exchange energy due to the spatially 
anisotropic character of the (unoccupied) Mo(4d) orbitals. For 
{\{}Mo$_{72}$Fe$_{30}${\}}, an angular range $103.6^\circ - 106.6^\circ$ is 
observed, which in part is caused by crystallographic disorder of Mo 
positions. To assess the influence of various geometric parameters involved 
in the superexchange pathways between two nearest-neighbor spin centers, 
both binding to a pentagonal diamagnetic 
[Mo$^{\mbox{\footnotesize{VI}}}_{6}$O$_{21}$(H$_{2}$O)$_{6}$]$^{6-}$ = 
{\{}Mo$_{6}${\}} 
fragment, we performed systematic Density Functional Theory-Broken Symmetry 
calculations\cite{22} on a model system in which two $s$ = 1/2 
[V$^{\mbox{\footnotesize{IV}}}$O(H$_{2}$O)$_{2}$]$^{2+}$ groups are coordinated to such a 
{\{}Mo$_{6}${\}} fragment in a nearest-neighbor (1,2) configuration. The 
fragment is augmented by an additional Zn$^{\mbox{\footnotesize{II}}}$(H$_{2}$O)$_{4}$ group 
binding in a (1,3,5) configuration for charge neutrality. The geometry of 
this model system was adjusted to match the actual configurations occurring 
in {\{}Mo$_{72}$Fe$_{30}${\}}. We find 
that $J$ in such a model system can deviate by up to $\pm $8{\%} from the 
average value $J_{0}$. Given the similarity between VO$^{2+}$, Cr$^{\mbox{\footnotesize{III}}}$, 
and Fe$^{\mbox{\footnotesize{III}}}$ in the Keplerate systems, namely that the magnetic orbitals 
cause a nearly isotropic spin density distribution, we expect that the 
variations in the relative values of $J$ span a very similar interval for 
{\{}Mo$_{72}$Fe$_{30}${\}} and {\{}Mo$_{72}$Cr$_{30}${\}}. As the 
intra-molecular variation in the values of the nearest-neighbor exchange 
constants implied by the optimal values (given in Sec.~\ref{sec:mul}) of the parameter 
$\rho$ are significantly larger, we suggest that this is due to 
numerous other perturbing mechanisms, some of which we listed above.

We also note that other attempts to explain limited features of $\partial 
M/\partial H$, specifically the broadening of the minimum versus $H$ for 
{\{}Mo$_{72}$Fe$_{30}${\}}, have been considered in the literature. One 
attempt assumed an elevated spin temperature during the pulsed field 
measurements, however this could be ruled out since a subsequent 
steady-field measurement reproduced the results obtained by the pulsed-field 
technique.\cite{17} Second, in a simulational study based on classical Monte 
Carlo calculations, effects of magnetic anisotropies, Dzyaloshinsky-Moriya, 
and dipole-dipole interactions have been considered.\cite{23} However, our own 
comprehensive simulational studies of these same mechanisms have shown that 
they give rise to only very minor corrections on the width of the minimum in 
$\partial M/\partial H$ versus $H$ for any reasonable choices of model 
parameters.

Finally, we suggest the additional possibility that in these molecules the variation of the exchange interaction is spontaneously generated so as to lower the system's magnetoelastic energy. Such behavior has been observed experimentally and described theoretically for a variety of antiferromagnetic oxide pyrochlore compounds\cite{25,26,27,28}. The pyrochlore lattice consists of corner-sharing tetrahedra and exhibits geometric frustration. In this regard one can understand this structure as the three-dimensional `cousin' of the corner-sharing triangle type structures realized in {\{}Mo$_{72}$Fe$_{30}${\}} and {\{}Mo$_{72}$Cr$_{30}${\}}.

In any event, it is highly satisfying that the 
frustrated magnetic molecules {\{}Mo$_{72}$Fe$_{30}${\}} and 
{\{}Mo$_{72}$Cr$_{30}${\}}, ostensibly zero-dimensional 
systems, are a source of novel and intriguing magnetic behavior. 
%%%%%%%%%%%%%%%%%%%%%%%%%%%%%%%%%%%%%%%%%%%%%%%%%%%%%%%%%%%%%%%%%%%%%%%%%%%%%%%%%%%%%%%%%%%%%%%%%
\begin{acknowledgments}
Research performed by C.S. at the Applied Sciences University Bielefeld was 
supported by an institutional grant. Work at the Ames Laboratory was 
supported by the Department of Energy-Basic Energy Sciences under Contract 
No. DE-AC02-07CH11358. R.P. acknowledges financial support from the Alfred 
P. Sloan Foundation. A.M. thanks the Deutsche Forschungsgemeinschaft, the Fonds der Chemischen Industrie, and the European Union for financial support. We thank H. Nojiri for sharing experimental data with us and for helpful discussions.
We also thank the thousands of volunteers participating in 
the public resource computing facility, Spinhenge@home 
[http://spin.fh-bielefeld.de]. The large-scale Monte Carlo simulations 
necessary for the present research were made possible due to the 
availability of their personal computers. 
\end{acknowledgments}
%%%%%%%%%%%%%%%%%%%%%%%%%%%%%%%%%%%%%%%%%%%%%%%%%%%%%%%%%%%%%%%%%%%%%%%%%%%%%%%%%%%%%%%%%%%%%%%%%

%%%%%%%%%%%%%%%%%%%%%%%%%%%%%%%%%%%%%%%%%%%%%%%%%%%%%%%%%%%%%%%%%%%%%%%%%%%%%%%%%%%%%%%%%%%%%%%%%


\begin{thebibliography}{9}
\bibitem{12} For classical spins systems that have frustrated ground states, the 
quantum analogs suffer from the so-called ``negative sign problem'' (see 
Ref.~\onlinecite{13}) and the quantum Monte Carlo method can provide reliable results 
only above a certain temperature whose value depends upon the details of the 
spin Hamiltonian (see Ref.~\onlinecite{14}).
\bibitem{16} By a classical isosceles spin triangle we mean a system of three 
classical spins (unit vectors $\vec {e}_i$) described by a Hamiltonian 
$J\left( {\vec {e}_1 \cdot \vec {e}_2 +\vec {e}_2 \cdot \vec {e}_3 } 
\right)+{J}'\,\vec {e}_3 \cdot \vec {e}_1 +\left( {\vec {e}_1 +\vec {e}_2 
+\vec {e}_3 } \right)\cdot \vec {H}$ (all quantities are dimensionless). If $J'=J$ we refer to the system as 
a classical equilateral spin triangle.
\end{thebibliography}
\end{document}